\documentclass[prd,reprint,aps,amsmath,amssymb,showpacs,nofootinbib]{revtex4-1}
\usepackage[utf8x]{inputenc}
\usepackage{graphicx}
\usepackage{enumitem}
\usepackage{slashed}
\usepackage{url}
\usepackage[pdftex]{hyperref}
\usepackage[T1]{fontenc}
\usepackage{bm}
\usepackage{color}
\usepackage{bbm}

\newcommand{\bra}[1]{\langle #1|}
\newcommand{\ket}[1]{|#1 \rangle}

\begin{document}

\title{$CP$ Violating Baryon Oscillations}
\preprint{INT-PUB-15-075}
\author{David McKeen}
\email{dmckeen@uw.edu}
\affiliation{Department of Physics, University of Washington, Seattle, WA 98195, USA}

\author{Ann E. Nelson}
\email{aenelson@uw.edu}
\affiliation{Department of Physics, University of Washington, Seattle, WA 98195, USA}                             
\date{\today}

\begin{abstract}
We enumerate the conditions necessary for $CP$ violation to be manifest in $n$-$\bar n$ oscillations, and build a simple model that can give rise to such effects. We discuss a possible connection between neutron oscillations and dark matter, provided the mass of the latter lies between $m_p-m_e$ and $m_p+m_e$.  We apply our results to   a possible  baryogenesis scenario  involving $CP$ violation in the oscillations of the $\Xi^0$.
\end{abstract}

\maketitle

\section{Introduction}
Given the evidence for an initial period of inflation in the early Universe, the asymmetry between matter and antimatter that we observe implies that baryon number ($B$) conservation must be violated~\cite{Sakharov:1967dj}. Furthermore, while the standard model (SM) accidentally conserves $B$ and lepton number ($L$) at the renormalizable level, both are broken by nonperturbative effects~\cite{Klinkhamer:1984di}, with only the combination $B-L$ conserved (however such anomalous processes cannot give rise to proton decay as they conserve $B$ mod 3).   If neutrino masses are of the Majorana type, as in the theoretically attractive seesaw mechanism,   they provide evidence for $B-L$ violation, via a $\Delta L=2$ operator, providing further motivation to consider $B$ violation. 

The two phenomenological hallmarks of $B$ violation are  neutron-antineutron oscillation~\cite{Kuzmin:1970nx,*Mohapatra:1980qe} and proton decay. Some recent reviews of $n$-$\bar n$ oscillation are given in Ref.~\cite{Babu:2013yww,*Babu:2013jba,*Phillips:2014fgb}. While proton decay is much more strongly constrained by experiment than neutron-antineutron oscillation, the former requires that $B$ is violated by one unit while the latter requires violation by two units. Therefore, in theories of baryon number violation that only admit $\Delta B=2$ operators, $n$-$\bar n$ oscillations can occur at a potentially observable rate without leading to proton decay (for recent work in this direction, see, e.g.~\cite{Mohapatra:2009wp,*Herrmann:2014fha,*Babu:2014tra,*Dev:2015uca,*Arnold:2012sd,Addazi:2015ata,*Addazi:2015goa}). For this reason, studying $n$-$\bar n$ oscillation is a critical component of understanding the origin of  $B$ violation.

The basic formulae governing neutron oscillation were first derived some time ago~\cite{Mohapatra:1980de,*Cowsik:1980np}. If $B$ is not conserved, there can be a nonzero transition amplitude between the (flavor eigenstates) $n$ and $\bar n$ that we denote as $\delta$. There is a splitting, $\Delta m$, between mass eigenstates, that are a linear combination of $n$ and $\bar n$, due to $\delta$ as well as from other interactions, e.g. with a background magnetic field.  Typically $\Delta m \gg \delta$. Using a basis in which the spin axis aligns with the background magnetic field, the system is effectively two independent two-state systems, with the component of angular momentum which is aligned with the magnetic field being conserved during the transition. For time scales short enough to neglect the possibility of neutron decay or when the neutron is stabilized by being in a stable nucleus,  the probability for a state that is a neutron at $t=0$ to oscillate to an antineutron is  \begin{align}
P_{n\to\bar n}=\frac{2\delta^2}{\Delta m^2}\left(1-\cos\Delta m\,t\right).
\end{align}
In a typical  experimentally relevant  situation where $\Delta m\,t\ll 1$, this becomes $P_{n\to\bar n}\simeq\delta^2t^2\equiv\left(t/\tau_{n\bar n}\right)^2$, which defines the oscillation lifetime. Neutron oscillations in nuclei lead to nuclear decays which presently set the best observational limit on $\tau_{n\bar n}$. The lower bound on the $^{16}{\rm O}$ lifetime from Super-Kamiokande of $1.9\times 10^{32}~\rm yr$~\cite{Abe:2011ky} at 90\% C. L. corresponds~\cite{Friedman:2008es} to $\tau_{n\bar n}>3.5\times10^8~\rm s$, or $\delta<1.9\times10^{-33}~\rm GeV$.     

In this letter we reconsider neutral baryon oscillations in the presence of   $CP$-violating and baryon number violating new physics, and discuss the conditions under which $CP$ violation can be exhibited in the oscillations. We also consider whether a dark matter particle could be observed in neutron decays, and whether such decays could contribute to observable $CP$ violation and/or baryogenesis.   We find that while  decays of oscillating neutrons are not  likely to exhibit a significant amount of $CP$ violation, it is possible  that   oscillations of neutral baryons containing heavier flavors could, perhaps even be enough to create the baryon asymmetry of the universe. We briefly outline a baryogenesis scenario involving $CP$ violation in oscillations of baryons containing heavy flavor.

\section{ CP  Violation in  neutral fermion oscillations}

Because  only states with the same spin can mix  (in a basis where the spin aligns with any external magnetic field), a two-state Hamiltonian $H$ suffices to describe oscillations of spin 1/2 particles. Here we use the 
$n$-$\bar n$ system as an example, but our results can be applied to any neutral spin 1/2 particle. In vacuum $H$ is given by
\begin{align}
 H &=\left( 
\begin{array}{cc} 
m_n-\frac{i}{2}\Gamma_n & M_{12}-\frac{i}{2}\Gamma_{12}    \\
 M^\ast_{12}-\frac{i}{2}\Gamma^\ast_{12}  &m_n-\frac{i}{2}\Gamma_n  \\
 \end{array}  \right).\\
\end{align}
For details of our formalism and how to derive $M_{12}$ and $\Gamma_{12}$  from a more fundamental theory follow from the treatment of Refs.~\cite{Ipek:2014moa,Ghalsasi:2015mxa} and are recapped in the Appendix.
While some authors~\cite{Berezhiani:2015uya} have recently claimed that $CP$ violation is required for neutron oscillations,   it is always possible to  reparameterize the system and use a definition of $CP$ so that the mass matrix is not $CP$ violating~\cite{Fujikawa:2015iia,Gardner:2016wov}. $CP$ violation in interference between mixing and decays is possible, leading to  the $n\to\bar n$ oscillation probability in vacuum differing from that for $\bar n\to n$,
\begin{align}
\frac{P_{\ket{n}\to\ket{\bar n}}}{P_{\ket{\bar n}\to\ket{n}}}-1=\frac{2\Im\left(M_{12}\Gamma_{12}^\ast\right)}{\left|M_{12}\right|^2-\left|\Gamma_{12}\right|^2/4-\Im\left(M_{12}\Gamma_{12}^\ast\right)}.
\end{align}
Generally, $\left|M_{12}\right|\gg\left|\Gamma_{12}\right|$ so that
\begin{align}
\frac{P_{\ket{n}\to\ket{\bar n}}}{P_{\ket{\bar n}\to\ket{n}}}-1\simeq\frac{2\left|\Gamma_{12}\right|}{\left|M_{12}\right|}\sin\beta,
\end{align}
with $\beta\equiv\arg M_{12}\Gamma_{12}^\ast$ a reparameterization-invariant $CP$-violating phase.  We now examine the characteristics necessary for a model that gives $n$-$\bar n$ oscillations at a rate not too far from the current upper bounds while also generating $CP$ violation that is not vanishingly tiny.

Models that violate baryon number only by two units allow for $n$-$\bar n$ oscillations without generating proton decay, which is subject to extremely strong constraints (see, e.g.,~\cite{Arnold:2012sd}). In such a model, if lepton number is not violated,  $\Gamma_{12}$ for the neutron system can be generated by operators that allow for the decays
\begin{align}
n\to\bar p e^+\nu_e,~\bar n\to p e^-\bar\nu_e
\end{align}
to proceed directly. However, the operators that generate these decays are dimension-12, and it is difficult for them to result in a value of $\Gamma_{12}$ that is not exceedingly small compared to $10^{-33}~\rm GeV$. We are therefore led to consider new states that are lighter than the neutron.

 As a motivation to consider such states,  let us  assume that baryon number is only absolutely conserved mod 2, so that  we have a conserved $\mathbb{Z}_2$ symmetry which is a subgroup of baryon number. This $\mathbb{Z}_2$ symmetry could be used to guarantee the stability of a dark matter Majorana fermion $\chi$, provided that $\chi$ is lighter than $m_p + m_e$. Stability of the proton requires than $\chi$ be heavier than $m_p - m_e$.\footnote{The stability of dark matter due to its mass being in this range in a model to explain baryogenesis was considered in Ref.~\cite{Allahverdi:2013mza}.} A slightly stronger lower bound of $m_\chi>937.9~\rm MeV$ comes from requiring that $^9{\rm Be}$ remain stable and not decay via the reaction $^9{\rm Be}\to{^8{\rm Be}}+\chi$. We have checked that this is the strongest bound that comes from requiring that all stable nuclides are kinematically forbidden from decaying to $\chi$ or $\chi e^+$. We then consider the decays of both neutron and antineutron into $\chi+\gamma$ in order to generate $\Gamma_{12}$. Justifying why $\chi$ should conveniently lie in this very narrow mass range is not the point of this letter, but  it could be argued for using anthropic reasoning, or perhaps $\chi$ could be a baryon of a sector mirror to ours~\cite{Berezhiani:2015afa,*Berezhiani:2005hv}, with its own gauge interactions. It is also possible that $\chi$  only couples to heavy flavors, but has a mass such that it can only decay into light flavors, with 2 or more  weak interactions are required for it to decay, in which case the constraint on its mass can be weakened.
 
As an example of how to generate an interaction that would allow  decays of the neutron or antineutron into $\chi+\gamma$, consider adding a scalar diquark, $\phi$, that is a color antifundamental and carries hypercharge $1/3$. Then it can couple to quarks through
\begin{align}
{\cal L}\supset g\phi\bar u \bar d+\rm h.c.,
\end{align}
where $\bar u$ and $\bar d$ are the (left-handed) up and down $SU(2)$ singlet quarks. We couple $\phi$ and $\bar d$ to the Majorana fermion $\chi$,
\begin{align}
{\cal L}\supset y\phi^\ast \bar d \chi+\rm h.c.
\end{align}

\begin{figure}
\includegraphics[width=0.7\linewidth]{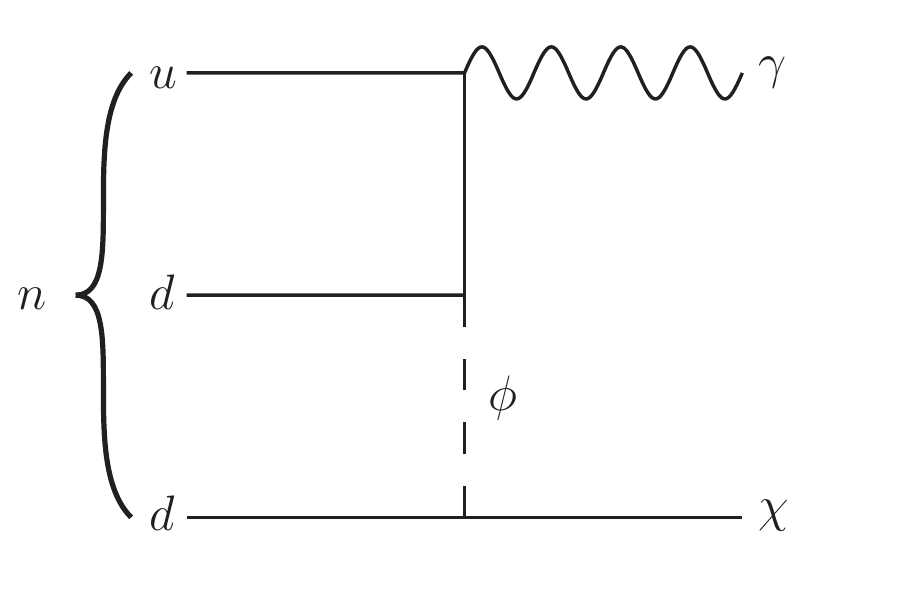}
\caption{One diagram that leads to the decay $n\to \chi\gamma$ as well as to an absorptive portion to the $n$-$\bar n$ transition amplitude, $\Gamma_{12}$, given a Majorana mass for $\chi$. Diagrams where the photon is attached to the scalar $\phi$ are suppressed by the large $\phi$ mass. Removing the photon gives the diagram responsible for $n$-$\chi$ (and $\bar n$-$\chi$) transitions.}\label{fig:ndecay}
\end{figure}
\begin{figure}
\includegraphics[width=\linewidth]{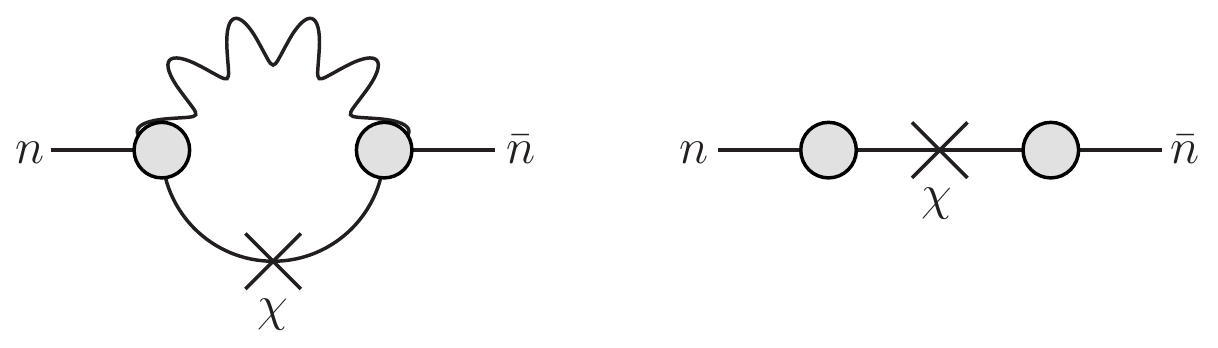}
\caption{Left: diagram responsible for $\Gamma_{12}$ through intermediate on-shell $\chi$ and $\gamma$. Right: diagram responsible for $M_{12}$. The crosses represent $\chi$ mass insertions and the blobs are the higher dimensional operators responsible for $n$-$\chi$ and $\bar n$-$\chi$ transitions, as seen in Fig.~\ref{fig:ndecay}.}\label{fig:G12M12}
\end{figure}

These interactions generate a transition dipole operator, through diagrams like the one shown in Fig.~\ref{fig:ndecay}, which appears as a term in the effective Lagrangian involving the neutron field as
\begin{align}
{\cal L}_{\rm eff}\supset \mu \bar\chi\Sigma_{\mu\nu}n F^{\mu\nu}+\rm h.c.,
\end{align}
with
\begin{align}
\mu\sim e \kappa\times\frac{g y}{m_\phi^2 m_n^2},
\end{align}
where $\kappa\sim 10^{-2}~\rm GeV^3$ comes from evaluating the hadronic matrix element $\bra{n}(udd)^2\ket{n}$ (for a detailed computation of this matrix element using lattice QCD, see~\cite{Buchoff:2015qwa}). As shown on the left of Fig.~\ref{fig:G12M12}, this coupling gives a contribution to $\Gamma_{12}$ that is
\begin{align}
\Gamma_{12}&=\frac{\mu^2 m_n^2 m_\chi}{16\pi}\left(1-\frac{m_\chi^2}{m_n^2}\right)^3
\\
&\sim 10^{-47}~{\rm GeV}\left(\frac{10^8~\rm GeV}{m_\phi/\sqrt{gy}}\right)^4\left(\frac{\Delta M}{1~\rm MeV}\right)^3.
\end{align}
where $\Delta M\equiv m_n-m_\chi\ll m_n$. There is also a contribution to $M_{12}$ from an off-shell intermediate $\chi$ exchange, seen on the right of Fig.~\ref{fig:G12M12}, that is larger than $\Gamma_{12}$,
\begin{align}
M_{12}&\sim\left(\frac{\kappa g y}{m_\phi^2}\right)^2\frac{m_\chi}{m_n^2-m_\chi^2}
\\
&\sim 10^{-33}~{\rm GeV}\left(\frac{10^8~\rm GeV}{m_\phi/\sqrt{gy}}\right)^4\left(\frac{1~\rm MeV}{\Delta M}\right).
\end{align}
For $CP$ violation to be present, there must exist another contribution to $M_{12}$, since this necessarily has the same phase as $\Gamma_{12}$. This can be done any number of ways, e.g.  we can add another Majorana fermion $\chi^\prime$  with similar couplings as $\chi$ but different phases in the couplings. Without fine-tuning the two separate contributions to $M_{12}$ against each other, we would then generically expect the level of $CP$ violation in the neutron-antineutron system to be tiny,
\begin{align}
\frac{P_{\ket{n}\to\ket{\bar n}}}{P_{\ket{\bar n}\to\ket{n}}}-1\propto\frac{\left|\Gamma_{12}\right|}{\left|M_{12}\right|}\lesssim10^{-14}\left(\frac{\Delta M}{1~\rm MeV}\right)^4.
\label{eq:asymm}
\end{align}

One may wonder whether there is a contribution to the neutron EDM that comes from attaching a photon to diagrams with external neutrons and a $\chi$-$\gamma$ loop. Since the amplitudes responsible for an EDM must have $\Delta B=0$ (as we argued above, and as also done in Ref.~\cite{Gardner:2016wov},)  potential contributions to the EDM from these diagrams must not contain a $B$-violating $\chi$ mass insertion. However, the lack of a factor of $m_\chi$ in such diagrams (in contrast to $M_{12}$ and $\Gamma_{12}$) means that they do not contain a physical $CP$-violating phase and thus do not contribute to the neutron EDM.

There is a contribution to the neutron EDM from a $d$ quark (chromo)EDM that arises through a $\chi$-$\phi$ loop. However, since we assume that the coupling of $\phi$ to quark doublets is absent, this requires a light quark mass insertion. Therefore $\phi$ masses in the $10-100~\rm TeV$ range are safe, even with $g\sim{\cal O}(1)$. For a detailed discussion of experimental limits on similar models, see~\cite{Arnold:2012sd,Berger:2010fy,*Baldes:2011mh}.

While a significant amount of $CP$ violation in $n$-$\bar n$ oscillations does not appear promising, this analysis suggests  that oscillations of neutral  baryons which are less directly constrained by nuclear stability can exhibit   larger $CP$ violation.  One might consider $\Lambda-\bar{\Lambda}$ oscillations~\cite{Kang:2009xt}, but $\Delta B=2$ operators with $\Delta S=2$ are highly constrained from decays of dinucleons into 2 kaons.  Other neutral baryons are more promising. Consider oscillations of $\Xi^0$ and $\bar\Xi^0$. Like neutron oscillation, this process involves operators that have $\Delta B=2$, but also requires $\Delta S=4$. Since four kaons are more massive than two nucleons, such operators are subject to   less stringent constraints than those with $\Delta S=0$ or $\Delta S=2$. In particular, they are not subject to strong limits from dinucleon decays.  It is   necessary to avoid generating a similar size for the much more constrained  baryon number violating $\Delta S=0, 1, 2$ and $\Delta S=3$~\cite{Glashow:2010vh} operators.  Provided that only SU(2) singlet strange quarks are involved in the   $\Delta S=4$  operators, then the radiative  generation of  $\Delta S=0,1$ and $\Delta S=2$ operators is suppressed by 2 more  weak loop factors as well as light quark mass insertions.  However a  $\Delta S=3$ operator may be generated via a single weak loop, greatly constraining the size of the operator which allows oscillations of the $\Xi^0$. Oscillations of baryons containing heavier flavors are less constrained. For instance   oscillations of the $\Xi^0_b$ were considered by Kuzmin in 1996~\cite{Kuzmin:1996sy} and  an oscillation rate which is comparable to the decay rate was shown    to be compatible with nuclear stability bounds.  Therefore, if $\phi$ couples dominantly to heavy flavor $SU(2)$ singlet quarks, it is conceivable that the strongest limits on $\phi$ comes from collider searches, and $m_\phi/\sqrt{gy}$ could be  a few hundred GeV or less. Detailed study of the various flavor combinations for neutral baryon oscillations is underway in Ref.~\cite{future}, and has shown that there are several baryons whose oscillation rates could be comparable to their decay rates without leading to excessive dinucleon decay.   In addition, if there exists a neutral $\chi$ fermion which both the baryon and anti baryon can decay into, the larger mass splitting between $\chi$ and the heavy baryon   allows for sizable  $\left|\Gamma_{12}/M_{12}\right]$. Given a mechanism for producing neutral heavy flavored baryons out of equilibrium in the early universe, such as via the late decay of some heavier $\chi$ particle as in Ref.~\cite{Ghalsasi:2015mxa},   then $CPV$ in their oscillations and decays is a potential    new mechanism for baryogenesis.  
\section{Conclusions}
There is firm evidence from the matter-antimatter asymmetry of the Universe that $B$ is violated.   Although proton decay searches are very sensitive probes of $B$ violation, they are less sensitive to theories that only allow for $\Delta B=2$ amplitudes. Searches for $n$-$\bar n$ oscillations directly test these theories. Much more rapid oscillations of heavy flavor baryons is allowed, and this possibility has not yet  been experimentally explored.  

In this letter we have enumerated the requirements for physical  $CP$ violation in neutral baryon oscillations. 
For the first time, a simple model that would give $CP$ violation in neutral baryon oscillations was built. The model connects the symmetry which suppresses proton decay to the stabilization of a dark matter candidate with mass around the mass of the proton.   The   amount of $CP$ violation in neutron oscillation in this model is still very small.   Heavy flavor baryon oscillations could  exhibit  significantly more $CP$ violation, and, given a mechanism (such as the one used in~\cite{Ghalsasi:2015mxa}) to produce them   in the early universe, after the QCD hadronization transition but before nucleosynthesis,   CP violation in heavy baryon oscillations and decays could be the origin of the matter-anti-matter asymmetry.  This possibility is currently under detailed study in Ref.~\cite{future}.

\appendix
 \section{}
Here we perform a detailed analysis of neutron oscillation in the generic case where the spin quantization axis does not align with the magnetic field, deriving the nonrelativistic Hamiltonian from the effective Lagrangian describing the neutron-antineutron system. Our aim is to clarify some statements in the literature and make our formalism more explicit.  We show that a simple two-state description of this phenomenon is sufficient\footnote{Electromagnetic and CPV effects were recently studied by  Gardner and Jafari in Ref.~\cite{Gardner:2014cma} and CP violation by Berezhiani and Vainshtein~\cite{Berezhiani:2015uya}. Our results are not in agreement with those published results, although Gardner and Yan have updated and extended her analysis~\cite{Gardner:2016wov}.}  (as previous studies~\cite{Mohapatra:1980de,*Cowsik:1980np} had used).   Similar work has been done recently \cite{Fujikawa:2015iia,Gardner:2016wov}, however as we consider  some new, previously unconsidered possibilities, such as exotic neutron decays and the possibility of observable $CP$ violation in neutron oscillations, for completeness we include our results here. 
\subsection{Deriving the Hamiltonian}
Our starting point is the four-component neutron field, $n$. It carries a charge $B=+1$ under a global $U(1)_B$ of baryon number. We write the most general Lagrangian density for describing a free neutron as
\begin{align}
{\cal L}&=\bar n\gamma^\mu \partial_\mu n+{\cal L}_{B}+{\cal L}_{\not B},
\end{align}
where the bar denotes Dirac conjugation, $\bar n=n^\dagger\gamma^0$. We have separated the kinetic term from the bilinear terms that conserve baryon number, ${\cal L}_{B}$, and those that violate baryon number, ${\cal L}_{\not B}$. The baryon-preserving terms are
\begin{align}
-{\cal L}_{B}&=\bar n\left(m_nP_L+m_n^\ast P_R\right)n,
\end{align}
where $P_{L,R}=(1\mp\gamma^5)/2$ project out the left and right chiralities of the four-component spinors. To construct the bilinears that break baryon number, we use the charge conjugated field $n^c$, which carries $B=-1$,
\begin{align}
-{\cal L}_{\not B}&=\bar{n}^c\left(\delta_1 P_L+\delta_2^\ast P_R\right)n+\bar {n}\left(\delta_2 P_L+\delta_1^\ast P_R\right)n^c.
\label{deltadef}
\end{align}

It can be useful to express the four-component spinors in the chiral basis in terms of left-handed, two-component Weyl spinors, $\xi$ and $\eta$ which carry $B=+1$ and $-1$ respectively,
\begin{align}
n=
\left(
\begin{array}{c}
\xi \\
\eta^\dagger 
\end{array}
\right),~
n^c=
\left(
\begin{array}{c}
\eta \\
\xi^\dagger 
\end{array}
\right).
\end{align}
Note that we suppress the spinor indices for clarity. For a thorough description of two-component spinor techniques, see~\cite{Dreiner:2008tw}. In terms of these fields,
\begin{align}
-{\cal L}_{B}&=m_n\eta\xi+{\rm h.c.},
\end{align}
and
\begin{align}
-{\cal L}_{\not B}&=\delta_1\xi\xi+\delta_2\eta\eta+{\rm h.c.}
\end{align}

We have defined charge conjugation so that
\begin{align}
n\underset{C}{\longrightarrow} n^c,~\xi\underset{C}{\longleftrightarrow} \eta.
\end{align}
Thus, $C$ leaves ${\cal L}_B$ unchanged for any $m_n$ and leaves ${\cal L}_{\not B}$ unchanged if $\delta_1=\delta_2$.

A parity transformation, $P$, flips helicities and can be implemented by
\begin{align}
n\underset{P}{\longrightarrow} \gamma^0n,~\xi\underset{P}{\longleftrightarrow} \eta^\dagger.
\end{align}
${\cal L}_B$ is therefore $P$-invariant if $m_n$ is real while ${\cal L}_{\not B}$ is parity invariant  if $\delta_1=\delta_2^\ast$.

Given the $C$ and $P$ transformations above, a combined $CP$ transformation (which is equivalent to time reversal $T$ since we are dealing with a local, Lorentz-invariant theory) takes
\begin{align}
n\underset{CP}{\longrightarrow} \gamma^0n^c,~\xi\underset{CP}{\longrightarrow} \xi^\dagger,~\eta\underset{CP}{\longrightarrow} \eta^\dagger.
\end{align}
$CP$ is then conserved by ${\cal L}_B$ and ${\cal L}_{\not B}$ if $m_n$, $\delta_1$, $\delta_2$ are all real.

To conserve $CP$, if $U(1)_B$ is violated, apparently requires removing three phases to make $m_n$, $\delta_1$, $\delta_2$ all real, while we only have two fields at our disposal to rephase, $\xi$ and $\eta$. Equivalently, in the four-component language, we can make a chiral transformation $n\to e^{i\alpha\gamma^5}n$ to make $m_n$ real and then $n\to e^{i \alpha'} n$ to remove the phase of $\delta_1$ or $\delta_2$, but not both. Thus at first glance  it appears that if $U(1)_B$ is broken, it is not possible to remove $CP$ violation. However, if $U(1)_B$ is only violated by $\delta_{1,2}$ then we can take a linear combination $n\to\cos\theta\, n+\sin\theta\, n^c$, $n^c\to\cos\theta\, n^c-\sin\theta\, n$, redefining $U(1)_B$, so that $\delta_1$ or $\delta_2$ are removed. Thus, $CP$ need not be violated by the mass terms since there are then only two phases to remove.   As long as we can redefine $CP$ to  the product of the original   $CP$ and any purely internal transformation, such that the new $CP$ is a symmetry,  there is no physically observable $CP$ violation. 

We also incorporate interactions with the electromagnetic field,
\begin{align}
{\cal L}_{\rm dipole}&=\frac{1}{2}F_{\mu\nu}\bar n\Sigma^{\mu\nu}\left(a P_L+a^\ast P_R\right)n
\\
&=a F_{\mu\nu}\eta\sigma^{\mu\nu}\xi+{\rm h.c.},
\end{align}
where $\Sigma^{\mu\nu}=(i/2)[\gamma^\mu,\gamma^\nu]$, and $\sigma^{\mu\nu}=(i/4)(\sigma^\mu\bar\sigma^\nu-\sigma^\nu\bar\sigma^\mu)$, $\sigma^\mu=(1,{\bm\sigma})$, $\bar\sigma^\mu=(1,-{\bm\sigma})$ with $\bm\sigma$ the Pauli matrices. $a=\mu_n-id_n$ with $\mu_n$, $d_n$ the neutron's magnetic (electric) dipole moment. We do not consider baryon-number--violating interactions with the electromagnetic field of the form $\bar{n}^c\Sigma^{\mu\nu} n\supset \xi\sigma^{\mu\nu}\xi$, $\eta\sigma^{\mu\nu}\eta$ since these vanish identically due to Fermi statistics.\footnote{In Ref.~\cite{Gardner:2014cma}, it was speculated that such interactions could develop through higher dimensional operators due to the composite nature of the neutron, as is the case for the neutron charge radius, however that work has been updated with different conclusions \cite{gardner}. In the language of effective field theory, a charge radius can develop because gauge invariance forbids writing the operator $\bar n \gamma^\mu A_\mu n$ (since the neutron is neutral) but allows $\bar n \gamma^\mu\partial^\nu F_{\mu\nu}n/\Lambda^2$ with $\Lambda$ the compositeness scale of the neutron. However, operators like $\bar{n}^c\Sigma^{\mu\nu} F_\mu^\rho(g_{\nu\rho}+\partial_\nu\partial_\rho/\Lambda^2+\dots)n$ couple two identical fermions to total angular momentum $J=1$. This configuration is symmetric under interchange of the identical fermion fields and therefore these operators vanish.}

To construct the Hamiltonian, we first introduce operators that create (and annihilate) neutron and antineutron states,
\begin{align}
\ket{n;{\bm p},s}={a_{\bm p}^s}^\dagger\ket{0},~\ket{\bar n;{\bm p},s}={b_{\bm p}^s}^\dagger\ket{0}.
\end{align}
The algebra satisfied by the creation and annihilation operators is
\begin{align}
\left\{a_{\bm p}^s,{a_{\bm k}^r}^\dagger\right\}=\left\{b_{\bm p}^s,{b_{\bm k}^r}^\dagger\right\}=\left(2\pi\right)^3\delta^{(3)}\left({\bm p}-{\bm k}\right)\delta^{sr},
\end{align}
with all other anticommutators zero. We decompose the Weyl spinors in terms of these as
\begin{align}
\xi_\alpha&=\sum_s\int\frac{d^3{\bm p}}{\left(2\pi\right)^{3/2}}\frac{1}{\sqrt{2E_{\bm p}}}\left[{x_{\bm p}^s}_\alpha a_{\bm p}^s e^{-ip\cdot x}+{y_{\bm p}^s}_\alpha {b_{\bm p}^s}^\dagger e^{ip\cdot x}\right],
\\
\eta_\alpha&=\sum_s\int\frac{d^3{\bm p}}{\left(2\pi\right)^{3/2}}\frac{1}{\sqrt{2E_{\bm p}}}\left[{x_{\bm p}^s}_\alpha b_{\bm p}^s e^{-ip\cdot x}+{y_{\bm p}^s}_\alpha {a_{\bm p}^s}^\dagger e^{ip\cdot x}\right],
\end{align}
where $E_{\bm p}=\sqrt{{\bm p}^2+m_n^2}$ if we assume that baryon number is nearly conserved, $\delta_{1,2}\ll m_n$. $x$ and $y$ are solutions to the Dirac equation and carry a spinor index. They can be expressed in terms of spin eigenstates $\omega^s$, such that ${\omega^s}^\dagger\omega^r=\delta^{sr}$, as
\begin{align}
&{x_{\bm p}^s}_\alpha=\sqrt{p\cdot\sigma}\omega^s, &{y_{\bm p}^s}_\alpha=2s\sqrt{p\cdot\sigma}\omega^{-s},
\\
&{x_{\bm p}^s}^\alpha=-2s{\omega^{-s}}^\dagger\sqrt{p\cdot\bar\sigma}, &{y_{\bm p}^s}_\alpha={\omega^s}^\dagger\sqrt{p\cdot\bar\sigma}.
\end{align}

The matrix elements of the Hamiltonian, $H^{ss^\prime}_{ij}$ ($i,j=n,\bar n$; $s,s^\prime=\pm 1/2$) for the neutron-antineutron system at rest can be computed through
\begin{align}
&-\bra{i;{\bm p},s}\int d^3x\,{\cal L}\ket{j;{\bm p}^\prime,s^\prime}\Big|_{{\bm p}\to0}
\\
&\quad\quad\quad\quad=\left(2\pi\right)^3\delta^{(3)}\left({\bm p}-{\bm p}^\prime\right)H^{ss^\prime}_{ij}.
\end{align}
To incorporate decays of the neutron and antineutron (as is necessary when discussing $CP$), we will decompose the Hamiltonian into a dispersive portion, $M$, and an absorptive part, $\Gamma$, generated by on-shell intermediate states,
\begin{align}
H=M-\frac i2\Gamma.
\end{align}

In the basis $(\ket{n,+},\ket{n,-},\ket{\bar n,+},\ket{\bar n,-})$, where the spin quantization axis is taken as the $\hat z$ direction, the dispersive portion of the Hamiltonian reads\footnote{Our Hamiltonian has $\bra{n;+}H\ket{\bar n;+}=\bra{n;-}H\ket{\bar n;-}$ while the one in Ref.~\cite{Gardner:2014cma} has $\bra{n;+}H\ket{\bar n;+}=-\bra{n;-}H\ket{\bar n;-}$ which violates Lorentz invariance. As pointed out in Ref.~\cite{Berezhiani:2015uya}, this error leads to incorrect eigenvalues of the Hamiltonian in Ref.~\cite{Gardner:2014cma}.}
\begin{align}
M&=\left( 
\begin{array}{cc}
\Re\left(m_n\right)\times\mathbbm{1}-{\bm H}\cdot{\bm\sigma}  & M_{12}\times\mathbbm{1}  \\
M_{12}^\ast\times\mathbbm{1} & \Re\left(m_n\right)\times\mathbbm{1}+{\bm H}\cdot{\bm\sigma}
\end{array}
\right),
\label{eq:M}
\end{align}
where ${\bm H}\equiv \mu_n {\bm B}-d_n {\bm E}$, $\mathbbm{1}$ is the $2\times2$ unit matrix, and $M_{12}\equiv \delta_1^\ast+\delta_2$.
While we have retained it in this expression for completeness, in what follows we ignore the (known to be tiny) electric dipole moment of the neutron and keep only the magnetic dipole moment. Because the dispersive portion of the Hamiltonian comes from on-shell intermediate states and we are assuming Lorentz invariance, $\Gamma$ must be trivial in spin. Thus, in this basis, it is given by
\begin{align}
\Gamma&=\left( 
\begin{array}{cccc}
\Gamma_n \times\mathbbm{1}  & \Gamma_{12}\times\mathbbm{1} \\
\Gamma_{12}^\ast\times\mathbbm{1} & \Gamma_n \times\mathbbm{1} 
\end{array}
\right),
\label{eq:Gamma}
\end{align}
where $\Gamma_n=1/885.6~{\rm s}=7.4\times10^{-28}~\rm GeV$ is the free neutron beta decay rate and $\Gamma_{12}$ represents any common final state that both neutron and antineutron may decay into (which necessarily requires physics beyond the SM). 

A $CP$ transformation on $H$ is implemented by taking the hermitian conjugate of $M$ and $\Gamma$ separately. Given the form of these matrices, we see that $CP$ violation requires a nontrivial phase difference between $M_{12}=\delta_1^\ast+\delta_2$ and $\Gamma_{12}$.

The Hamiltonian in this basis can be broken up into $2\times2$ blocks according to its effect on baryon number,
\begin{align}
H&\sim\left( 
\begin{array}{cc}
\Delta B=0 & \Delta B=2 \\
\Delta B=-2 & \Delta B=0
\end{array}
\right).
\end{align}
The $\Delta B=0$ blocks mix spins (if there is an external electromagnetic field) while, because of Fermi statistics, the $\Delta B=\pm 2$ blocks do not. Because of this, even in the presence of  (constant) electromagnetic fields, a $2\times2$ Hamiltonian is sufficient to describe the neutron-antineutron system. To see that, we make a unitary transformation of the Hamiltonian using the matrix
\begin{align}
U_B&=\left( 
\begin{array}{cccc}
c_\theta & e^{-i\alpha}s_\theta & 0 & 0 \\
0 & 0 & c_\theta & e^{-i\alpha}s_\theta \\
-e^{i\alpha}s_\theta & c_\theta & 0 & 0 \\
0 & 0 & -e^{i\alpha}s_\theta & c_\theta
\end{array}
\right),
\end{align}
with $\tan 2\theta=\sqrt{B_x^2+B_y^2}/B_z$ and $\tan\alpha=B_y/B_x$. Operating on $H$ with this matrix gives
\begin{widetext}
\begin{align}
U_BHU_B^\dagger&=\left( 
\begin{array}{cccc}
m_n-\mu_n B_z-\frac i2\Gamma_n  & H_{12} & 0 & 0 \\
H_{21} & m_n+\mu_n B_z-\frac i2\Gamma_n & 0 & 0 \\
0 & 0 & m_n+\mu_n B_z-\frac i2\Gamma_n  & H_{12} \\
0 & 0  & H_{12} & m_n-\mu_n B_z-\frac i2\Gamma_n
\end{array}
\right),
\end{align}
\end{widetext}
defining $H_{12}\equiv M_{12}-(i/2)\Gamma_{12}$, $H_{21}\equiv M_{12}^\ast-(i/2)\Gamma_{12}^\ast$. Thus we see that the system described by the $4\times4$ Hamiltonian given by Eqs.~(\ref{eq:M}) and (\ref{eq:Gamma}) is actually two separate, identical two-state systems---this transformation aligned the spin quantization axis along the direction of the magnetic field. This justifies previous analyses of $n$-$\bar n$ transitions that used a two-state Hamiltonian~\cite{Mohapatra:1980de,*Cowsik:1980np}, despite the presence of electromagnetic fields that could have required the consideration of neutron and antineutron spin.\footnote{The time evolution is more complicated if the direction of the external field is time dependent, as considered in Ref.~\cite{Gardner:2014cma}, as the matrix which block diagonalizes the Hamiltonian is then time dependent. However,   the neutron-antineutron  mixing  angle is always suppressed by an external magnetic field. We do not consider time varying fields here.}

To diagonalize $H$ we make use of the matrix $U$,
\begin{widetext}
\begin{align}
U&=\left( 
\begin{array}{cccc}
H_{12}^{-1/2}c_1c_\theta & H_{12}^{-1/2}c_1e^{-i\alpha}s_\theta & -H_{21}^{-1/2}s_1c_\theta & -H_{21}^{-1/2}s_1e^{-i\alpha}s_\theta \\
-H_{12}^{-1/2}c_1e^{i\alpha}s_\theta & H_{12}^{-1/2}c_1c_\theta & -H_{21}^{-1/2}s_1e^{i\alpha}s_\theta & H_{21}^{-1/2}s_1c_\theta \\
H_{12}^{-1/2}s_2c_\theta & H_{12}^{-1/2}s_2e^{-i\alpha}s_\theta & H_{21}^{-1/2}c_2c_\theta & H_{21}^{-1/2}c_2e^{-i\alpha}s_\theta \\
H_{12}^{-1/2}s_2e^{i\alpha}s_\theta & -H_{12}^{-1/2}s_2c_\theta &  -H_{21}^{-1/2}c_2e^{i\alpha}s_\theta & H_{21}^{-1/2}c_2c_\theta
\end{array}
\right),
\end{align}
which gives
\begin{align}
UH U^{-1}&=\left( 
\begin{array}{cccc}
m_n-\Delta/2-\frac i2\Gamma_n  &  &  &  \\
 & m_n+\Delta/2-\frac i2\Gamma_n  &  &  \\
 &  & m_n+\Delta/2-\frac i2\Gamma_n  &  \\
  &  &  & m_n-\Delta/2-\frac i2\Gamma_n
\end{array}
\right).
\end{align}
\end{widetext}
In these expressions, we have taken $m_n$ real, defined $\Delta\equiv2\sqrt{\mu_n^2 B^2+H_{12}H_{21}}$, and we use $s_\theta$ and $c_\theta$ to denote $\sin\theta$ and $\cos\theta$ with $\theta$ as given above. We have also defined $c_{1,2}=N_{1,2}\sqrt{1+z}$, $s_{1,2}=N_{1,2}\sqrt{1-z}$ with $z=2\mu_n B/\Delta$ and $N_{1,2}^2=\left|H_{12}\left(1\pm z\right)\right|+\left|H_{21}\left(1\mp z\right)\right|$.

\subsection{Transition Probabilities}
The transition probabilities are given in terms of the matrix $U$ by
\begin{align}
P_{i\to j}=\left|U_{ki}U^{-1}_{jk}e^{-iM_kt}\right|^2,
\end{align}
with $M_K$ the mass of eigenstate $k$. As mentioned, because the $\Delta S=\pm 2$ amplitudes do not mix spins, the probability for a neutron to oscillate into an antineutron of the opposite spin (or vice versa) is computed to be zero,
\begin{align}
P_{\ket{n;+}\to\ket{\bar n;-}}&=P_{\ket{\bar n;+}\to\ket{n;-}}=0.
\end{align}
We find the probability as a function of time for a neutron at $t=0$ to transition to an antineutron of the same spin (and omitting superfluous spin indices) to be
\begin{align}
P_{\ket{n}\to\ket{\bar n}}&=\frac{2\left|H_{21}\right|^2}{\left|\Delta\right|^2}\left(\cosh\frac{\Delta\Gamma t}{2}-\cos\Delta mt\right)e^{-\Gamma_n t},
\label{eq:Pnnbar}
\end{align}
and that for an antineutron at $t=0$ to oscillate to a neutron is
\begin{align}
P_{\ket{\bar n}\to\ket{n}}&=\frac{2\left|H_{12}\right|^2}{\left|\Delta\right|^2}\left(\cosh\frac{\Delta\Gamma t}{2}-\cos\Delta mt\right)e^{-\Gamma_n t}.
\label{eq:Pnbarn}
\end{align}
We note here that, for $\Gamma_{12}=0$, our expressions for the oscillation probabilities in Eqs.~(\ref{eq:Pnnbar}) and (\ref{eq:Pnbarn}) agree with standard formulae that have appeared in the literature~\cite{Mohapatra:1980de,*Cowsik:1980np} and, in particular, are suppressed when $\mu_nB\gg\left|H_{12}\right|$, $\left|H_{21}\right|$. The current experimental limit on the $^{16}{\rm O}$ lifetime translates to $\left|H_{12,21}\right|\lesssim 10^{-33}~\rm GeV$.

\begin{acknowledgments}
We thank Patrick Draper, Marat Freytsis, Susan Gardner, Seyda Ipek,  Maxim Pospelov, and Arkady Vainshtein for helpful discussions. We  thank the Institute of Nuclear Theory  and the organizers of the INT program {\it Intersections of BSM Phenomenology and QCD for New Physics Searches}. This work was supported in part by the U.S. Department of Energy under Grant No. DE-FG02-96ER40956.
\end{acknowledgments}

\bibliography{nnbar}

\end{document}